\documentclass[twocolumn,showpacs,preprintnumbers,amsmath,amssymb,prb,floatfix]{revtex4}

\usepackage{epsfig,psfrag}
\usepackage{color}
\usepackage{graphicx}% Include figure files
\usepackage{dcolumn}% Align table columns on decimal point
\usepackage{bm}% bold math
\def \rd{\mathrm{d}}

\begin{document}

\title{Finite-size version of the excitonic instability in graphene     
quantum dots}

\author{Tomi Paananen and Reinhold Egger}
\affiliation{ Institut f\"ur Theoretische Physik, Heinrich-Heine-Universit\"at, 
D-40225 D\"usseldorf, Germany }

\date{\today}
\begin{abstract}
By a combination of Hartree-Fock simulations, exact diagonalization, and
perturbative calculations, we investigate the ground-state properties of 
disorder-free circular quantum dots formed in a graphene monolayer. 
Taking the reference chemical potential at the Dirac point, 
we study $N\le 15$ interacting particles, where the  
fine structure constant $\alpha$ parametrizes the Coulomb interaction. 
We explore three different models: (i) Sucher's positive
projection (``no-pair'') approach, (ii) a more general Hamiltonian 
conserving both $N$ and the number of additional 
electron-hole pairs, and (iii) the full quantum 
electrodynamics (QED) problem, where only $N$ is conserved.  
We find that electron-hole pair production is important for $\alpha\agt 1$. 
This corresponds to a reconstruction of the filled Dirac sea and 
is a finite-size version of the bulk excitonic instability.
We also address the effects of an orbital magnetic field.
\end{abstract}


\maketitle

\section{Introduction}
\label{sec1}

Coulomb interaction effects in monolayer graphene\cite{geim,castro}  
are currently attracting a lot of attention; for a recent review, 
see Ref.~\onlinecite{kotov}.  From a theory point
of view, this interest mainly stems from the possibility of realizing
a strong-coupling version of QED in a readily accessible 
two-dimensional (2D) system.  In fact, the (bare) fine structure constant is 
rather large, $\alpha=e^2/(\hbar\kappa v_F)\simeq 2.2/\kappa$, 
with the effective substrate dielectric constant $\kappa$ and
the Fermi velocity $v_F\approx c/300 \approx 10^6$~m$/$s.
Retardation effects are irrelevant here, i.e., we effectively have
2D massless Dirac fermions interacting via the Coulomb potential. 
Similar physics can be expected for the surface state in 3D topological
insulators,\cite{hasan} but interactions are expected to be much
weaker due to the large $\kappa$ in the relevant materials.
In graphene, the situation away from the Dirac point 
(defined as zero of energy) can be reasonably well 
understood in terms of Fermi liquid theory,\cite{kotov,gonzalez} 
but the picture is more complicated near the Dirac point.  
At a critical interaction strength $\alpha_c$,
a semimetal-insulator transition is theoretically expected\cite{khvec}
due to electron-hole proliferation.  For $\alpha>\alpha_c$, a finite gap
corresponding to an excitonic insulator is formed and the
ground state undergoes reconstruction. On the other hand,
quantum critical behavior is expected\cite{son,schmalian} as a precursor 
of the instability for $\alpha<\alpha_c$.
Recent lattice quantum Monte Carlo simulations\cite{drut} found
the critical value $\alpha_c\approx 1.1$ for an infinitely extended (``bulk'') 
graphene monolayer.  Similar values were also obtained analytically from the 
dynamical polarization function approach\cite{gamayun} and
under the ladder approximation to the Bethe-Salpeter equation.\cite{fertig}
However, so far no experimental signature of this \textit{excitonic
instability} has been reported.
It has also been recognized that the excitonic instability for the 
bulk many-body problem is related to the simpler ``supercritical'' 
instability of the hydrogen problem in graphene,\cite{kotov}  where
$\alpha$ corresponds to the (attractive) potential strength of the nucleus.
Above a critical value for $\alpha$, the nucleus captures an electron 
to screen its positive charge below criticality,
while at the same time a hole escapes to infinity in order to maintain charge
neutrality.  In atomic physics, essentially the same phenomenon
should also take place for superheavy atoms.\cite{greiner}
The creation of an electron-hole pair thus also
accompanies the supercritical instability.
In the presence of an homogeneous orbital magnetic field $B$, the hole
escape process is disturbed by the formation of closed Landau orbits.
For the bulk many-body problem, the resulting 
magnetic catalysis phenomenon\cite{gusy} implies a lowering of 
$\alpha_c$ with increasing $B$. 

In this work, we study a finite-size version of the
excitonic instability presumably realized in available graphene quantum dots.
Quantum dots in conventional 2D systems have been studied 
extensively,\cite{kouwenhoven,reimann} and experimental results 
for lithographically prepared graphene dots were reported
recently.\cite{exp1,exp2,exp3,exp4,exp5}  Within the single-particle picture, 
theoretical proposals on how to model such a dot have been reviewed 
in Ref.~\onlinecite{castro}. We here adopt the probably simplest route by
imposing the so-called ``infinite-mass boundary condition,''\cite{berry,schnez} 
where no current is allowed to flow through the circle $r=R$ defining the
 dot's boundary.  While disorder limits the quality of the boundary in existing
dots,\cite{exp5} such a boundary condition captures at 
least their qualitative physics.  Moreover, future experimental progress
 is likely to yield well-defined boundaries.

We investigate the ground state of $N$ interacting electrons
in  a  closed circular graphene dot, where $N$ particles are added on top of
the filled Dirac sea, i.e., relative to the chemical potential $\mu=0$.
This problem has been studied before within the Hartree-Fock (HF) 
approach.\cite{hf1,hf2,jpa,tomi} However, when going  
beyond effective single-particle theory, one has to deal with the 
``Brown-Ravenhall disease,''\cite{brown,sucher} i.e., 
the possibility to excite electron-hole pairs with small
energy by combining a hole and an electron both very far away from the 
Fermi surface.   While such processes are physically suppressed 
by the finite bandwidth, the infinitely deep filled Dirac sea present 
in the Dirac theory renders naive approaches mathematically ill-defined.   
For $\alpha\ll 1$ and when a gap separates 
electron and hole states, Sucher\cite{sucher} showed that one can 
circumvent the Brown-Ravenhall problem 
by a suitable projection $\Lambda_+$ of the basic QED
Hamiltonian $H$ to a well-defined no-pair Hamiltonian 
$H_+=\Lambda_+ H\Lambda_+$, where the filled Dirac sea is 
effectively treated as completely inert.
The projection operator $\Lambda_+$ eliminates 
negative-energy (hole) states from the single-particle Hilbert space.
We study the validity of the no-pair approach for graphene
dots and find that for $\alpha\alt 1$, it is indeed meaningful,
see also Ref.~\onlinecite{haus1}. 
On a quantitative level, however, it is accurate only for $\alpha\ll 1$. 
As also discussed by Sucher,\cite{sucher} if one wishes to go beyond 
the positive projection scheme,  a QED approach is indicated. The 
QED Hamiltonian $H$, see Eq.~\eqref{qed} below,
does not conserve the number $N_{eh}$ of electron-hole pairs. 
In fact, only the particle number $N$ -- defined as the 
imbalance of electron and hole numbers -- is conserved, and 
a superposition of states with different $N_{eh}$  
determines the ground state for strong interactions. Once
electron-hole pairs proliferate, a reconstruction of the ground state 
takes place.  We encounter this phenomenon for $\alpha\agt 1$
in graphene dots, similar to the reported critical value\cite{drut}
for the bulk excitonic instability.  However, in our finite-size system  this
is a smooth crossover and not a phase transition.
We stress that our ``$N$-particle problem'' defines
$N$ as the difference of electron and hole numbers, 
which allows for the excitation of an arbitrary number 
$N_{eh}$ of electron-hole pairs. 
For $\alpha\to 0$, this definition reduces to having $N$ electrons on
top of the filled Dirac sea.

The structure of the remainder of this paper is as follows.
In Sec.~\ref{sec2} we introduce the model and discuss the various
theoretical approaches employed to find the ground state.
An intermediate approach is to generalize the no-pair
 approach (where $N_{eh}=0$) 
to allow for a fixed but finite number $N_{eh}$ of electron-hole pairs.  
The Hamiltonian $H_{\rm fix}$ is obtained from $H$ by neglecting all
 terms that do not conserve $N_{eh}$.  A sufficient (but not 
necessary) condition for the breakdown of the no-pair 
Hamiltonian $H_+$  arises when the ground-state energy
of $H_{\rm fix}$ is lowered for some $N_{eh}>0$.
We assess the validity of the no-pair scheme
in Sec.~\ref{sec3} by comparing to results obtained 
under $H_{\rm fix}$ and from the QED Hamiltonian $H$.  We
perform these calculations using exact diagonalization (ED) 
for $N=2$ and $N=3$ particles in the dot. 
In Sec.~\ref{sec4}, we use $H_+$ to carry out
detailed HF calculations for up to $N=15$ particles
and relatively weak interactions, $\alpha\le 1$. We present
results for the ground-state spin, valley
polarization, and addition energy as function of $N$.
Finally, in Sec.~\ref{sec5} we provide a discussion of our main results.

\section{Model and theoretical approaches}
\label{sec2}

In this section, we describe the model employed in our study of
the electronic properties of interacting graphene quantum dots.
We will then turn to different theoretical approaches 
to obtain the ground-state properties.

\subsection{Single-particle problem}

It is well established that on low energy scales, quasiparticles in graphene
are described by the Dirac Hamiltonian,\cite{castro}
\begin{equation}\label{dw}
H_0 = v_F {\bm \sigma}\cdot \left( {\bf p}+\frac{e}{c}{\bf A}\right) + 
M(r)\sigma_z \tau_z - \mu_B {\bf s}\cdot {\bf B},
\end{equation}
where ${\bf p}=-i\hbar (\partial_x,\partial_y)^T$ 
and $-e$ is the electronic charge. 
The Pauli matrices ${\bm \sigma}=(\sigma_x,\sigma_y)$ and $\sigma_z$
refer to graphene's sublattice structure, while the
Pauli matrix $\tau_z$ corresponds to the valley degree of freedom, i.e., to
the two $K$ points. 
 A static vector potential ${\bf A}({\bf r})$ [with ${\bf r}=(x,y)^T$] 
 allows for the inclusion of a constant orbital magnetic field $B_z$,
where we choose the symmetric gauge, ${\bf A}=\frac12 B_z (-y,x)^T$.
Since we neglect spin-orbit couplings in Eq.~\eqref{dw}, spin
Pauli matrices ${\bf s}=(s_x,s_y,s_z)$ only appear in the Zeeman 
term. With $\mu_B$ denoting Bohr's magneton and putting the Land{\'e}
factor to $g_e=2$,\cite{exp5} this term couples to the full
(homogeneous) magnetic field, ${\bf B}=(B_x,B_y,B_z)$ with $B=|{\bf B}|$.
Switching to polar coordinates $(r,\phi)$,  
we consider a clean circular quantum dot in a graphene monolayer modelled
by the well-known infinite-mass boundary condition,\cite{berry}
where the mass $M(r)$ in Eq.~\eqref{dw} is zero 
for $r<R$ but tends to $+\infty$ for $r>R$. This choice
ensures that no current flows through the boundary at $r=R$. 
Eigenstates can be classified by the conserved 
total angular momentum, $J_z=-i \hbar\partial_\phi+\hbar\sigma_z/2$, 
with eigenvalue $\hbar j$ for half-integer $j=m+1/2, m\in \mathbb{Z}$.

While the eigenfunctions can be found in analytical form even in the
presence of the magnetic field,\cite{schnez} the Coulomb interaction
matrix elements are readily available\cite{tomi} only in the $B=0$ basis.
We therefore first describe the solution for $B=0$ and later 
include the homogeneous magnetic field.
Note that different valleys ($\tau=\pm$) are 
decoupled and spin ($s=\pm$) then simply yields a twofold degeneracy. 
For given $(m,\tau,s)$, we first discuss the $E>0$ 
solutions to $H_0 \Phi^{(+)} =E\Phi^{(+)}$, 
where the spinor has the sublattice structure
\begin{equation}
\Phi^{(+)}(r,\phi)= e^{im\phi}\left(\begin{array}{c}
\psi_{1,m}(r) \\ ie^{i\phi} \psi_{2,m}(r) \end{array}\right).
\end{equation}
The infinite-mass boundary condition implies\cite{berry,schnez}
\begin{equation}\label{nbc}
\psi_{1,m}(R) = \tau \psi_{2,m}(R).
\end{equation}
With the Bessel functions $J_m(kr)$ of the first kind,
$k= E/\hbar v_F$, and normalization constant $A$,
the Dirac equation for $r<R$ is solved by the \textit{Ansatz} 
\[
\psi_{1,m}(r)=AJ_m(kr),\quad \psi_{2,m}=A J_{m+1}(kr).
\]
The quantization condition \eqref{nbc} then determines the
 eigenenergies $E_a>0$ with $a\equiv (n,m,\tau,s)$,
\begin{equation}\label{ener}
J_m( E_{a}/\Delta_0 ) = \tau J_{m+1}( E_{a}/\Delta_0 ),
\end{equation}
where $\Delta_0\equiv \hbar v_F/R$ is the single-particle level spacing 
of the dot and $n\in \mathbb{N}$ labels different solutions
for given $(m,\tau,s)$. Equation (\ref{ener}) is easily solved numerically
and the eigenstates to energy $E_a>0$ are
\begin{equation}
\Phi_{a}^{(+)}(r,\phi) = A_a e^{im\phi} \left(\begin{array}{c}
J_m ( k_a  r) \\ i e^{i\phi} J_{m+1}(k_a r) \end{array}\right),
\end{equation}
where $k_a=E_a / \hbar v_F$ and the normalization factor is
\begin{equation}\label{amn}
A_{a} = [\pi ( J_m^2 - J_{m-1}J_{m+1} + J_{m+1}^2 - J_{m} J_{m+2})]^{-1/2}
\end{equation}
with $J_{m}\equiv J_m(E_a/\Delta_0)$.  Time-reversal invariance implies the 
Kramers degeneracy relation $E_{n,m,\tau,s}=E_{n,-m-1,-\tau,-s}$.
Negative-energy (hole) solutions, $\Phi_{\tilde a}^{(-)}(r,\phi)$, 
follow by using the electron-hole symmetry property
of the Hamiltonian, $E_{n,m,-\tau,s}=-E_{n,m,\tau,s}$. 
We use the multi-index $a$ ($\tilde a$) to count states with
positive (negative) energy.  There is no zero-energy solution 
for $B=0$, and we have a finite gap around the Dirac point.  

Next we add the magnetic field. 
Expressed in terms of the eigenstates $\Phi^{(+)}_{a}$ 
and $\Phi^{(-)}_{\tilde a}$, 
the vector potential part in $H_0$ has a matrix structure diagonal both in
the quantum numbers $(m, \tau, s)$ and the conduction/valence band index 
$\pm$, i.e., only different $n$  states are mixed.  By numerical
diagonalization, it is straightforward to obtain the resulting eigenenergies
 $\tilde E_a>0$ and $\tilde E_{\tilde a}<0$, and the 
corresponding eigenstates. The indices $n$ and thus $a$ ($\tilde a$) are
redefined to take into account the unitary transformation diagonalizing 
$H_0$.
Finally, we include the Zeeman term.
Choosing the spin quantization axis along ${\bf B}$, 
where $s=\pm 1$ corresponds to spin-up or spin-down states, the  
full eigenenergy $E_a>0$ is given by\cite{foot1}
\begin{equation} 
E_{a} =  \tilde E_{a} - s \mu_B B,
\end{equation}
and similary for $E_{\tilde a}<0$. In a slight abuse of notation,
$E_a$ now denotes the full eigenenergy and not the solution to 
Eq.~\eqref{ener} anymore.
The Zeeman term is generally quite small\cite{castro} but
breaks the spin degeneracy of the levels, while
the vector potential breaks the valley degeneracy, see Eq.~\eqref{nbc}.
The resulting eigenstates are 
denoted by $\Phi^{(+)}_{a}(r,\phi)$ and $\Phi^{(-)}_{\tilde a}(r,\phi)$.

\subsection{Many-body interactions}

We now include the Coulomb interaction among the 
particles. The noninteracting reference problem is characterized
by a filled Dirac sea ($\mu=0$), i.e., all $E_{\tilde a}<0$ states are filled.
The QED Hamiltonian $H$ describing this
problem can be expressed in terms of electron annhilation 
operators, $c_a$, corresponding to the single-particle states 
$\Phi_a^{(+)}$, and hole creation operators, $d_{\tilde a}^\dagger$, 
with single-particle states $\Phi_{\tilde a}^{(-)}$.
The full field operator is written as
\begin{equation}\label{field}
\Psi({\bf r}) = \sum_a \Phi_a^{(+)}({\bf r}) c_a + \sum_{\tilde a}
\Phi^{(-)}_{\tilde a}({\bf r}) d_{\tilde a}^\dagger.
\end{equation}

Since the Hamiltonian commutes with $\tau_z$ and ${\bf s}\cdot {\bf B}$, 
we can write $\Psi({\bf r}) = \sum_{\tau s} \Psi_{\tau s}({\bf r})$.
The Hamiltonian is then given by $H=H_k+H_I$,
with the kinetic part (note that $E_{\tilde a}<0$)
\begin{equation}
H_k= \sum_a E_a c_a^\dagger c^{}_a +
\sum_{\tilde a} |E_{\tilde a}| d^\dagger_{\tilde a} d_{\tilde a}^{}
\end{equation}
and the interaction term
\begin{eqnarray}\label{hi}
H_I&=&\frac{\hbar v_F\alpha}{2} \sum_{\tau\tau'ss'}
\int \frac{ \rd {\bf r} \rd {\bf r}'}{|{\bf r}-{\bf r}'|} 
\\ \nonumber &\times&
: \Psi^\dagger_{\tau s} ({\bf r}) \Psi^\dagger_{\tau' s'} ({\bf r}') 
\Psi^{}_{\tau' s'} ({\bf r}') \Psi^{}_{\tau s} ({\bf r}) : ,
\end{eqnarray}
where the colons denote normal ordering.
Inserting the field operator expansion (\ref{field}) into Eq.~\eqref{hi},
\begin{equation}\label{qed}
H=H_{\rm fix}+H'.
\end{equation}
$H_{\rm fix}$ commutes separately with both the electron 
and the hole number operator, $\hat N_e=\sum_a c^\dagger_a c^{}_a$ and
$\hat N_h=\sum_{\tilde a} d^\dagger_{\tilde a} d_{\tilde a}^{}$.  
The full Hamiltonian, however, only commutes with $\hat N=\hat N_e-\hat N_h$. 
We thus define the $N$-particle problem  
by having $N$ excess electrons and $N_{eh}=N_h$ electron-hole pairs on top
of the filled Dirac sea.  Only $N$ is conserved, while $N_{eh}$ can fluctuate. 
Under $H_{\rm fix}$ alone, the number $N_{eh}$ of 
electron-hole pairs is conserved,
\begin{eqnarray} \nonumber H_{\rm fix} &=& H_k+ 
\frac12 \sum_{aba'b'} \left(V_{abb'a'}-\delta_{ss'}
V_{aba'b'}\right) c_a^\dagger c_b^\dagger c_{b'}^{} c_{a'}^{} \\
\label{hnp} &+& \frac12 \sum_{\tilde a \tilde b \tilde a' \tilde b'}
 \left(V_{\tilde a\tilde b \tilde b' \tilde a'}-\delta_{ss'}
V_{\tilde a \tilde b \tilde a' \tilde b'}\right) 
d_{\tilde a}^\dagger d_{\tilde b}^\dagger d_{\tilde b'}^{} d_{\tilde a'}^{} \\
\nonumber &-&
\sum_{aa',\tilde b\tilde b'} 
\left( V_{a\tilde b\tilde b' a'}-\delta_{ss'} V_{a\tilde b a' \tilde b'}
\right) c^\dagger_a d^\dagger_{\tilde b} d_{\tilde b'}^{} c_{a'}^{}.
\end{eqnarray}
All terms not commuting with $\hat N_{e,h}$ are 
collected in the remaining part, $H'= h+h^\dagger$, with
\begin{eqnarray}\label{hprime}
h &=&\frac12 \sum_{ab\tilde a' \tilde b'} 
\left( V_{ab \tilde b' \tilde a'}-\delta_{ss'}
V_{ab\tilde a'\tilde b'}\right) c_a^\dagger c_b^\dagger d_{\tilde
b'}^{\dagger} d_{\tilde a'}^{\dagger}  
\\ &+& \nonumber \frac12 \sum_{a a' \tilde b\tilde b' }  
\delta_{s,-s'} V_{a\tilde b \tilde a' b'}
c_a^\dagger d_{\tilde a'}^\dagger d_{\tilde b}^{} c_{b'}^{}  
\\ &+& \nonumber \sum_{abb'\tilde a'}  \left( V_{ab b'\tilde a'}-\delta_{ss'}
V_{bab'\tilde a'}\right) c_a^\dagger d_{\tilde a'}^\dagger 
c_b^{\dagger} c^{}_{b'}  \\ &-& \nonumber 
\sum_{a\tilde a' \tilde b\tilde b'}  
\left( V_{a\tilde b\tilde b'\tilde a'}-\delta_{ss'}
V_{a\tilde b\tilde a'\tilde b'}\right) c_a^\dagger d_{\tilde a'}^\dagger 
d_{\tilde b'}^{\dagger} d^{}_{\tilde b}.
\end{eqnarray}
In Eqs.~\eqref{hnp} and \eqref{hprime},
the spin quantum numbers are given by $s=s_a=s_{a'}$ and $s'= s_b=s_{b'}$
(when hole states are involved, $a\to \tilde a$ etc.).
These spin selection rules  are encoded in the
interaction matrix elements $V_{aa'b'b}$ which have been derived in a 
form useful for numerical evaluation in Ref.~\onlinecite{jpa}. We
quote them for the convenience of the reader next.  
A finite matrix element follows only when the valley selection rule, 
$\tau_{a}=\tau_{a'}$ and $\tau_b=\tau_{b'}$, and 
angular momentum conservation, $m_a+m_{a'}=m_b+m_{b'}$, are satisfied.
When all selection rules are met,
\begin{widetext}
\begin{eqnarray*}
V_{aa'b'b} &=& (4\pi)^2 \alpha \Delta_0 
A_{a} A_{a'}A_{b'} A_b \sum_{l=0}^\infty
C_{q,l} \int_0^1 \rd r \ r^{-l} \left(J_{m_a}(E_a r) J_{m_b}(E_br) +
J_{m_a+1}(E_a r) J_{m_b+1}(E_b r)\right)\\ 
&\times & \int_0^r \rd r' \ (r')^{l+1} \left(
J_{m_{a'}}(E_{a'} r') J_{m_{b'}}(E_{b'}r' )
+J_{m_{a'}+1}(E_{a'} r') J_{m_{b'}+1}(E_{b'}r' ) \right) ,
\end{eqnarray*}
\end{widetext}
with $E_a$ in units of $\Delta_0=\hbar v_F/R$ and $q\equiv |m_b-m_a|$.
The coefficient $C_{q,l}$ vanishes when $l+q$ is odd or when $l<q$.
For $q=l=0$, we have $C_{0,0}=1/2$, while otherwise 
\[
C_{q,l} = \frac{(2l-1)!!}{2^{l+1} l!} \prod_{n=1}^{(l+q)/2}
\frac{(n-1/2)(n-l-1)}{n(n-l-1/2)}.
\]

\subsection{Calculation approaches}\label{sec2c}

A standard way to proceed is to employ the no-pair approach.\cite{tomi} 
With the projector $\Lambda_+$ to 
the subsector $E_a>0$ of the single-particle Hilbert space (for each particle),
we thus consider the $N$-particle problem with respect to the
filled Dirac sea.  The projected Hamiltonian 
$H_+=\Lambda_+ H \Lambda_+=\Lambda_+ H_{\rm fix} \Lambda_+$ is
given by
\begin{eqnarray}\label{hplus}
H_+ &=& \sum_a E_a c_a^\dagger c^{}_a \\ \nonumber
&+& \frac12 \sum_{aba'b'} \left(V_{abb'a'}-\delta_{ss'}
V_{aba'b'}\right) c_a^\dagger c_b^\dagger c_{b'}^{} c_{a'}^{} .
\end{eqnarray}
As detailed in Refs.~\onlinecite{jpa,tomi}, this 
allows for a straightforward implementation of the HF
 approach, and we will report HF results in Sec.~\ref{sec4}. In contrast
to Refs.~\onlinecite{jpa,tomi}, we here include the valley and spin
degrees of freedom. 
Given the converged self-consistent density matrix, one can obtain
the ground-state energy $E(N)$, the total spin quantum number $S$
of the $N$-electron dot from the eigenvalues $\hbar^2 S(S+1)$ 
of $\sum_{i=1}^N {\bf S}_i^2$, and the valley polarization eigenvalue 
$\tau(N)=\sum_i \tau_i$.

On a more general level, we allow for a fixed number of electron-hole
pairs by employing the Hamiltonian $H_{\rm fix}$ [Eq.~\eqref{hnp}].
The no-pair Hamiltonian $H_+$ follows from
$H_{\rm fix}$ with $N_{eh}=0$.  When the ground-state
energy of $H_{\rm fix}$ is minimized for some $N_{eh}>0$, the no-pair
approach breaks down. Interactions are then able to overcome the 
gap between valence and conduction band, and one cannot treat the Dirac sea
as inert anymore.  $H_{\rm fix}$ as well as the QED model $H$ are
studied by ED and perturbation theory in Sec.~\ref{sec3}.

\section{Particle-hole pair production and reconstruction of the ground state}
\label{sec3}

\begin{figure}[t]
\begin{center}
\includegraphics[width=3in]{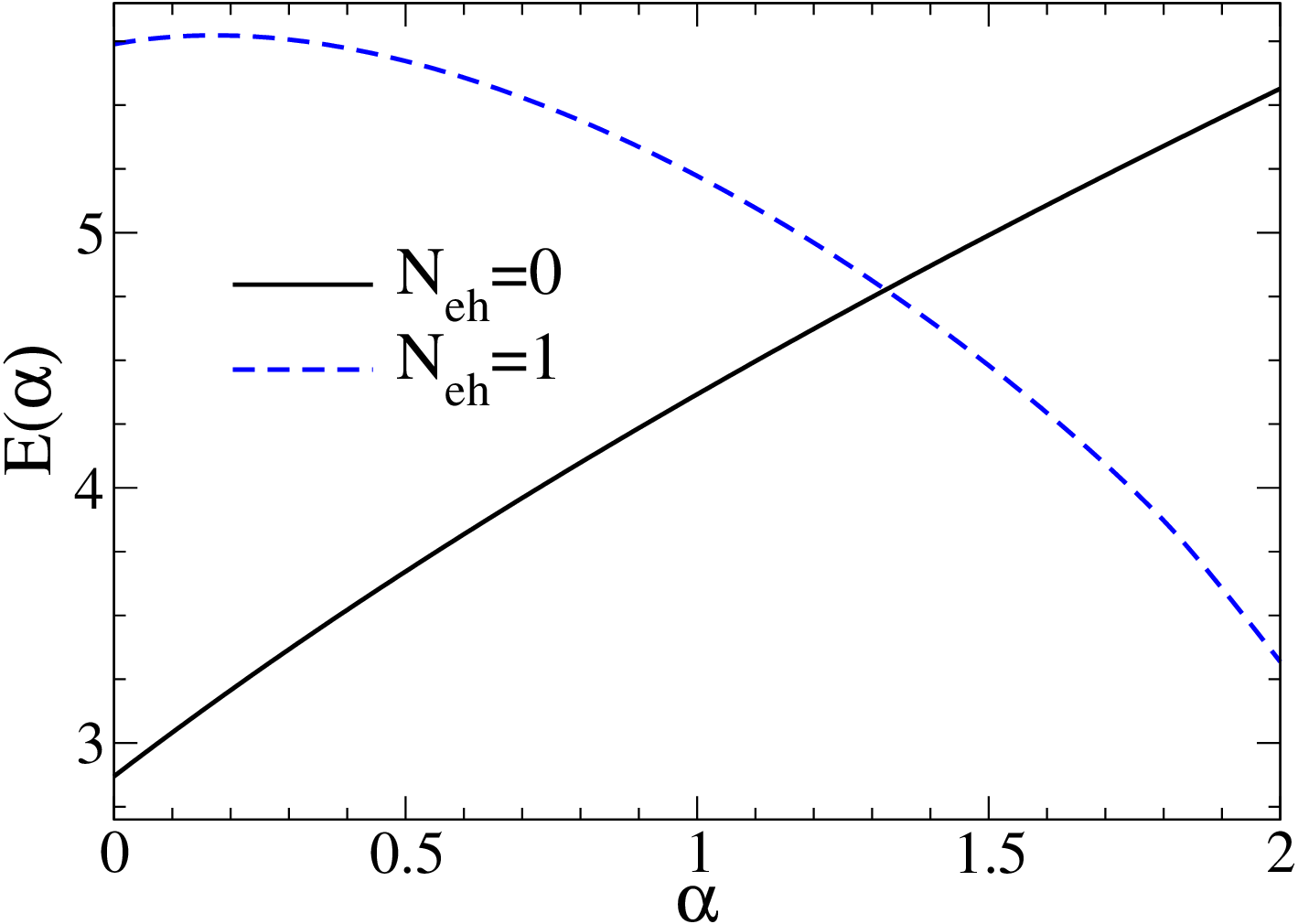}
\caption{\label{fig1} (Color online)
ED results for the ground-state energy $E$ in units of $\Delta_0=\hbar v_F/R$
 vs fine structure constant  $\alpha$
for $N=2$ particles in a graphene dot with $B=0$. We here  use the
Hamiltonian $H_{\rm fix}$ in Eq.~\eqref{hnp}.\\ }
\end{center}
\end{figure}
\begin{figure}
\begin{center}
\includegraphics[width=3in]{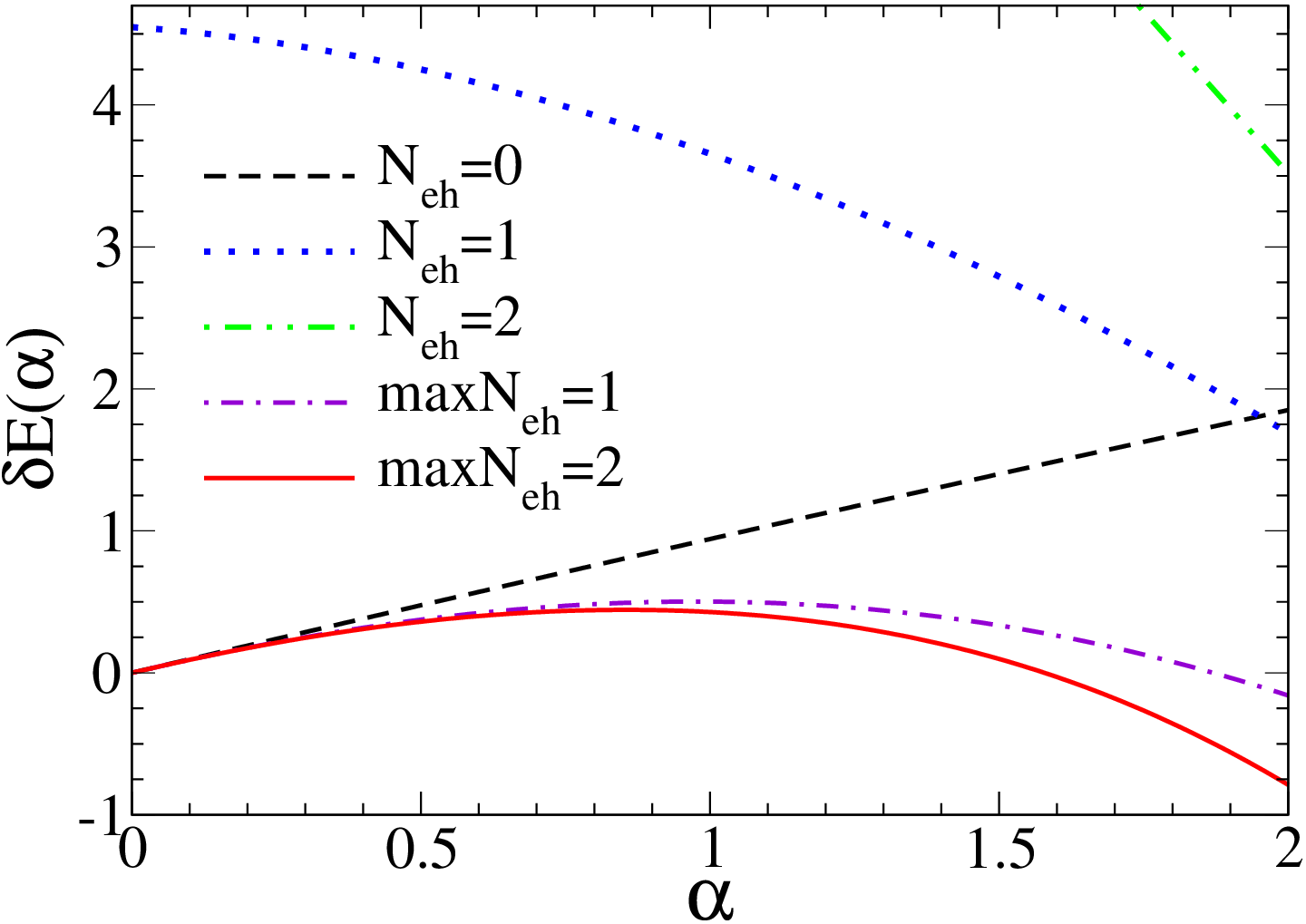}
\caption{\label{fig2} (Color online)
ED results for the interaction energy  $\delta E(\alpha)=E(\alpha)-E(0)$ 
(in units of $\Delta_0$) vs $\alpha$ for $N=2$. We consider
a spinless single-valley version of graphene with $B=0$. 
The curves for $N_{eh}=0,1,2$ correspond to the
Hamiltonian $H_{\rm fix}$ with $N_{eh}$ electron-hole pairs, i.e., 
the ground state then has no electron-hole pair for $\alpha\alt 1.9$.
However, the full QED Hamiltonian \eqref{qed}, where we truncate the 
Hilbert space to at most one or two electron-hole pairs (${\rm max}(N_{eh})
=1,2$), has a significantly lower energy already for $\alpha\agt 0.5$. \\}
\end{center}
\end{figure}
\begin{figure}
\begin{center}
\includegraphics[width=3in]{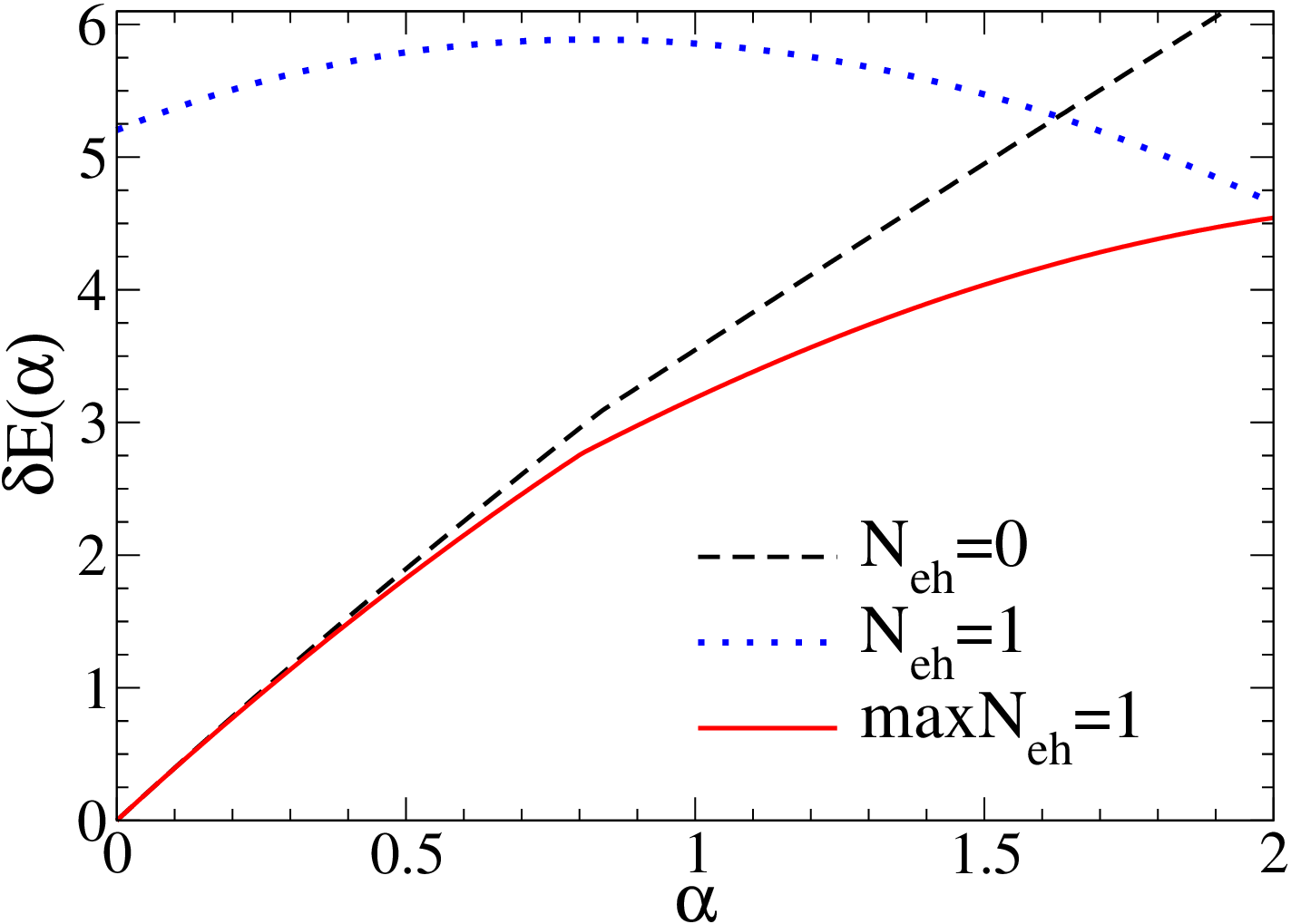}
\caption{\label{fig3} (Color online)
Same as Fig.~\ref{fig2} but for $N=3$.\\}
\end{center}
\end{figure}

We now compare the three theoretical approaches
in Sec.~\ref{sec2c} by employing ED for particle numbers $N=2$ and $3$.
Convergence in the ED calculations was achieved by keeping about
30 single-particle states (per spin and valley degree of freedom), 
and memory size limitations represented the main bottleneck. 
For a given $\alpha$, ED results can be obtained within a few minutes
on a standard desktop computer.

Figure \ref{fig1} shows results for the $\alpha$-dependence of the 
ground-state energy for $N=2$ using $H_{\rm fix}$ with $N_{eh}=0$ and $1$.  
We observe that for $\alpha\alt 1.3$, the ground state of $H_{\rm fix}$
contains no electron-hole pair, but for stronger interaction,
the ground state undergoes reconstruction and involves at least
one electron-hole pair.  The no-pair Hamiltonian $H_+$
thus necessarily fails when $\alpha\agt 1.3$.  Moreover, as we shall discuss
next, the presence of $H'$ [Eq.~\eqref{hprime}] restricts its applicability 
even further.

Since ED of the QED Hamiltonian $H$ in Eq.~\eqref{qed}
is computationally very demanding even for $N=2$, in the remainder of this
section, we shall restrict ourselves to a spinless single-valley 
version of graphene.  The ED results obtained from $H_{\rm fix}$ and $H$ are
compared for $N=2$ in Fig.~\ref{fig2}. 
For the spinless single-valley version of $H_{\rm fix}$,
no electron-hole pairs are excited in the ground state for $\alpha\alt 1.9$.  
However, the $H'$ contribution is important already for $\alpha\agt 0.5$, 
 see Fig.~\ref{fig2}. The full interaction
correction to the energy is significantly lowered by 
including $H'$ and may even change sign for large $\alpha$.  
In these calculations, the Hilbert
space was truncated to contain at most one or two electron-hole pairs.
For $\alpha\alt 1.5$, this appears to be sufficient.  Figure \ref{fig3}
shows results for $N=3$, where we arrive at similar conclusions. 

\begin{figure}[t]
\begin{center}
\includegraphics[width=3in]{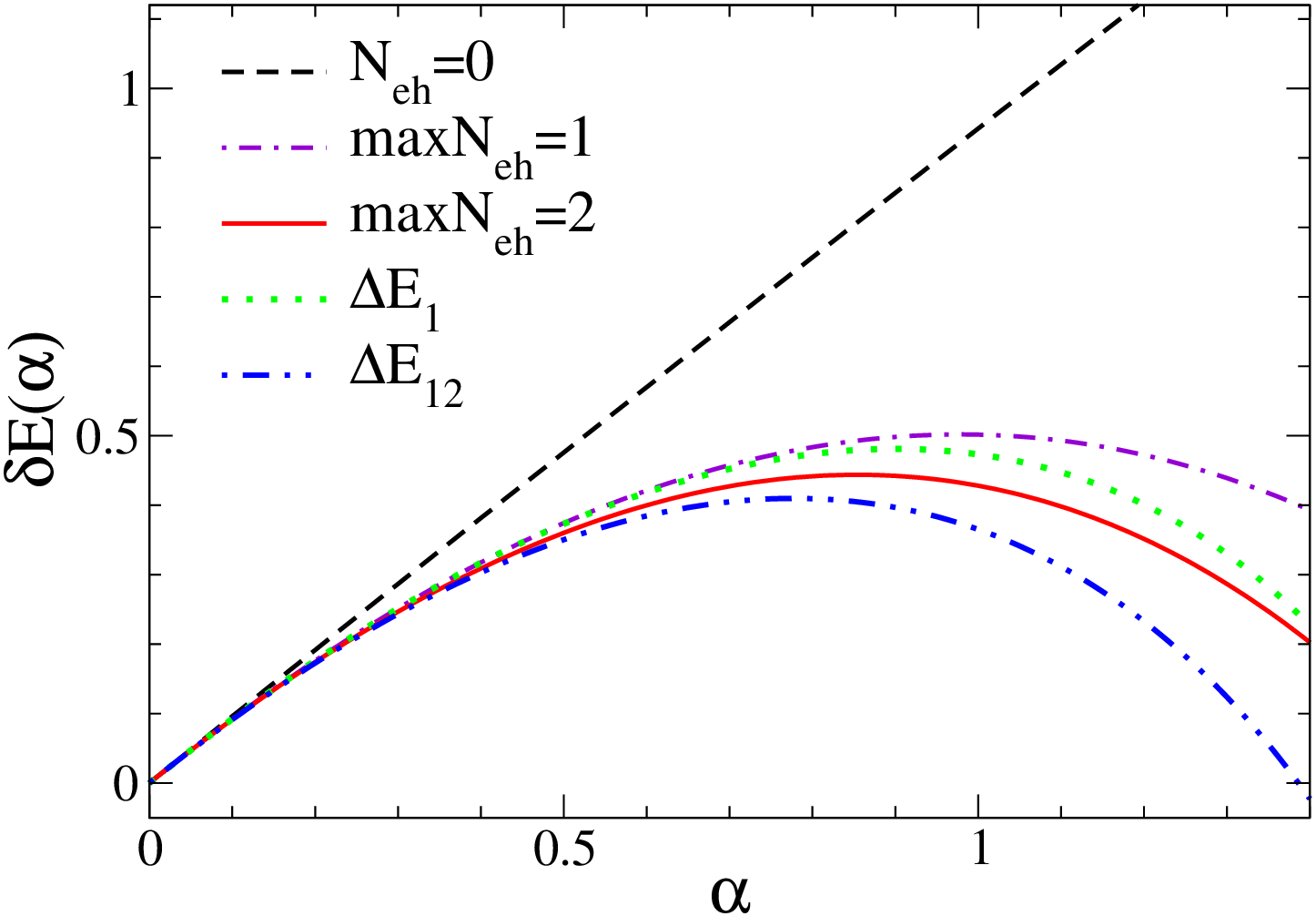}
\caption{\label{fig4} (Color online)
Same as Fig.~\ref{fig2} but including the results
of second-order perturbation theory in $H'$.
 The curve $\Delta E_1$ takes into account corrections involving 
one electron-hole pair only, while $\Delta E_{12}$ is the full second-order
result including up to two pairs.\\ }
\end{center}
\end{figure}
\begin{figure}
\begin{center}
\includegraphics[width=3in]{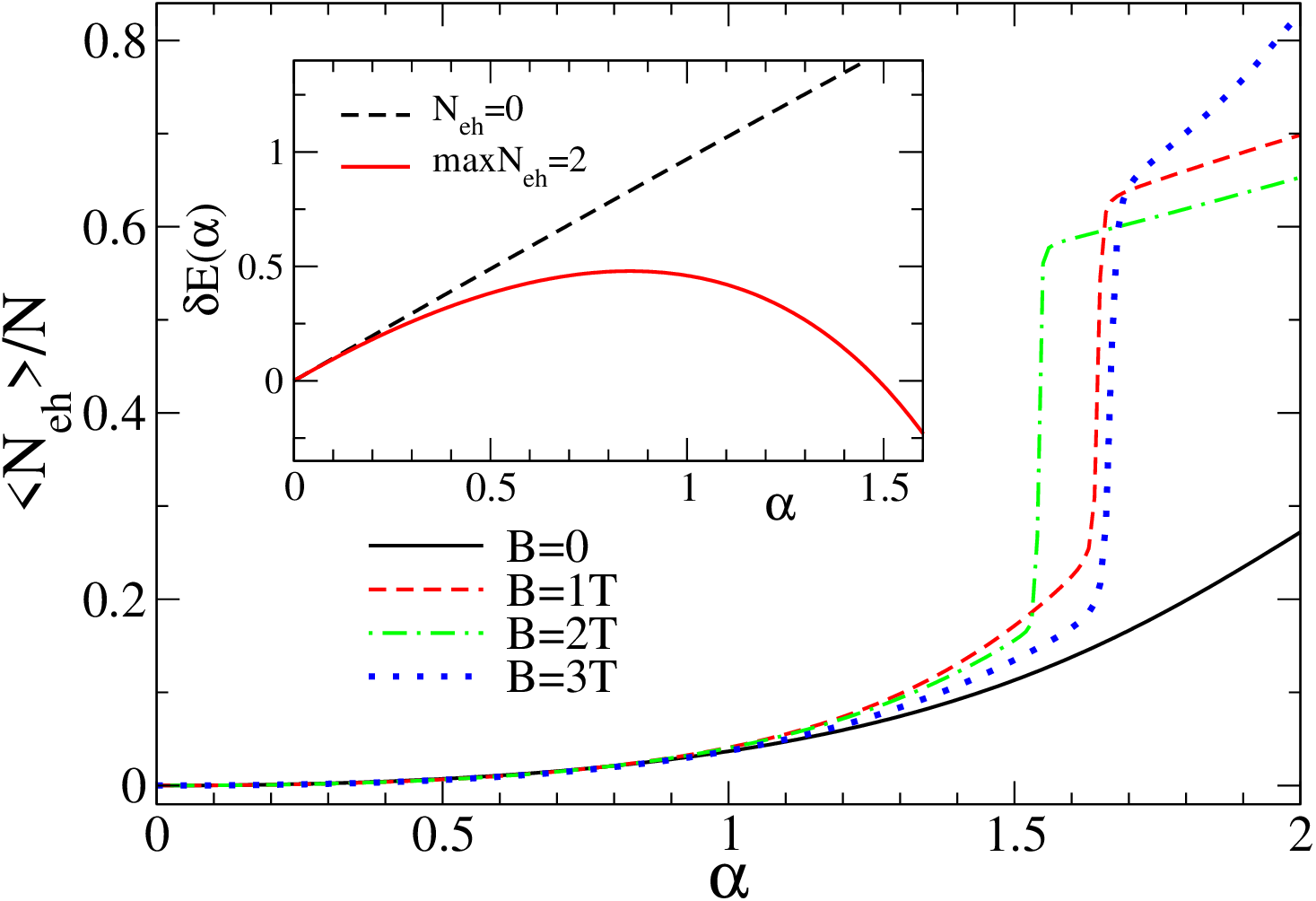}
\caption{\label{fig5} (Color online)
Main panel: Relative number of electron-hole pairs
in the ground state, $\langle N_{eh}\rangle/N$, 
vs $\alpha$ for $N=2$ and several values of 
the magnetic field.  We take the dot radius $R=30$~nm.
The results were obtained using the QED Hamiltonian
but with the Hilbert space truncated to at most two electron-hole pairs.
Inset: Interaction correction $\delta E$ vs $\alpha$ for $N=2$ and $B=1$~T.
Shown are ED results using $H_+$ ($N_{eh}=0$) and for the full $H$, 
where the Hilbert space was truncated at ${\rm max} (N_{eh})=2$.\\ }
\end{center}
\end{figure}

The effect of $H'$ can also be evaluated analytically by using second-order 
perturbation theory (the first order vanishes identically). The
result is shown for $N=2$ in Fig.~\ref{fig4}, together with the
ED results from Fig.~\ref{fig2}.  
We see that second-order perturbation theory captures the ED
 data quite well, especially for $\alpha\alt 1$.  
The same conclusion was reached for $N=3$ (results not shown here),
and the combination of ED (or HF) calculations
for $H_{\rm fix}$ supplemented with a perturbative treatment
of $H'$ should in general provide a good approximation of the ground state. 

Let us now discuss the case of finite magnetic
field, again for the computationally simpler spinless single-valley case 
with $N=2$. We have also studied $N=3$ particles,
again with very similar results. 
The main panel of Fig.~\ref{fig5} shows the average number of
electron-hole pair excitations in the ground state for
several values of the magnetic field.  The shown results are for 
a dot radius $R=30$~nm. A magnetic field of $B=1$~T corresponds to the
magnetic length $l_B=(c/eB)^{1/2}\approx 26$~nm, which is of the
same order of magnitude as the radius.
Evidently, in a magnetic field, the proliferation of electron-hole
pairs becomes more important. We interpret this effect as the finite-size
analogue of the magnetic catalysis phenomenon.\cite{gusy}
The interaction correction to the ground-state energy is shown
for $B=1$~T in the inset
of Fig.~\ref{fig5}. While the result shows qualitatively similar behavior 
as for $B=0$, the now more significant deviations between the ED data and the 
no-pair result are consistent with magnetic catalysis again. 
We note in passing that for $\alpha\agt 1.5$, the basis size used in our
ED calculations is most likely not sufficient, and
probably $N_{eh}>2$ states also contribute to the ground state.
The steplike features in Fig.~\ref{fig5} are then presumably 
smeared out.

We conclude that the no-pair Hamiltonian is quantitatively
reliable only for weak interactions, $\alpha\alt 0.5$, and
for not too large magnetic fields.  For stronger interactions and/or fields, 
the ground state undergoes reconstruction and electron-hole pair
proliferation.  Using $H_{\rm fix}$ [Eq.~\eqref{hnp}] is not sufficient to 
get more accurate results, and one has to include $H'$ 
[Eq.~\eqref{hprime}] which does \textit{not}\ conserve the electron and
hole numbers separately.  However, for $\alpha\alt 1$, 
quite accurate results for the ground-state
energy are obtained by combining ED (or HF) calculations for
the no-pair Hamiltonian with subsequent second-order perturbation theory
 in $H'$.   Here only two or three particles have been addressed, 
where electron-hole proliferation takes place around $\alpha\approx 1$.  
Since the bulk case, which follows from the above model by a suitable 
limiting procedure with $N\to \infty$ and $R\to \infty$, 
has a phase transition  at $\alpha\approx 1.1$, the 
finite-size crossover apparently depends on $N$ only weakly.

\section{Addition spectrum and ground-state properties}
\label{sec4}

\begin{figure}[t]
\begin{center}
\includegraphics[width=3in]{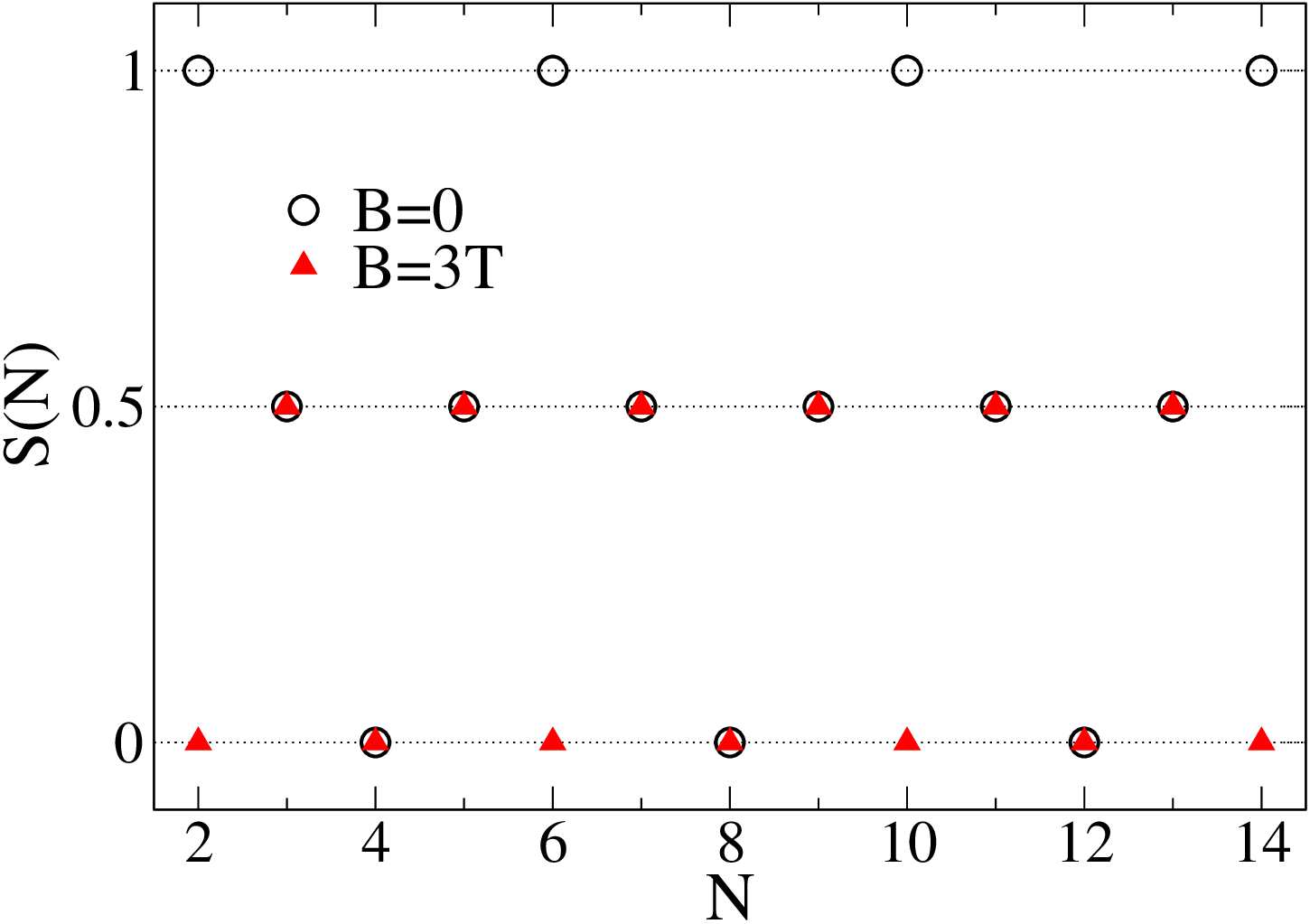}
\caption{\label{fig6} (Color online)
HF results using the no-pair model, $H_+$, for the 
ground-state spin $S$ vs particle number $N$ for  
$B=0$ and for $B=B_z=3$~T (with $R=30$~nm).  
For all shown $N$ and $\alpha\le 1$, 
$S(N)$ does not depend on $\alpha$. \\ }
\end{center}
\end{figure}
\begin{figure}
\begin{center}
\includegraphics[width=3in]{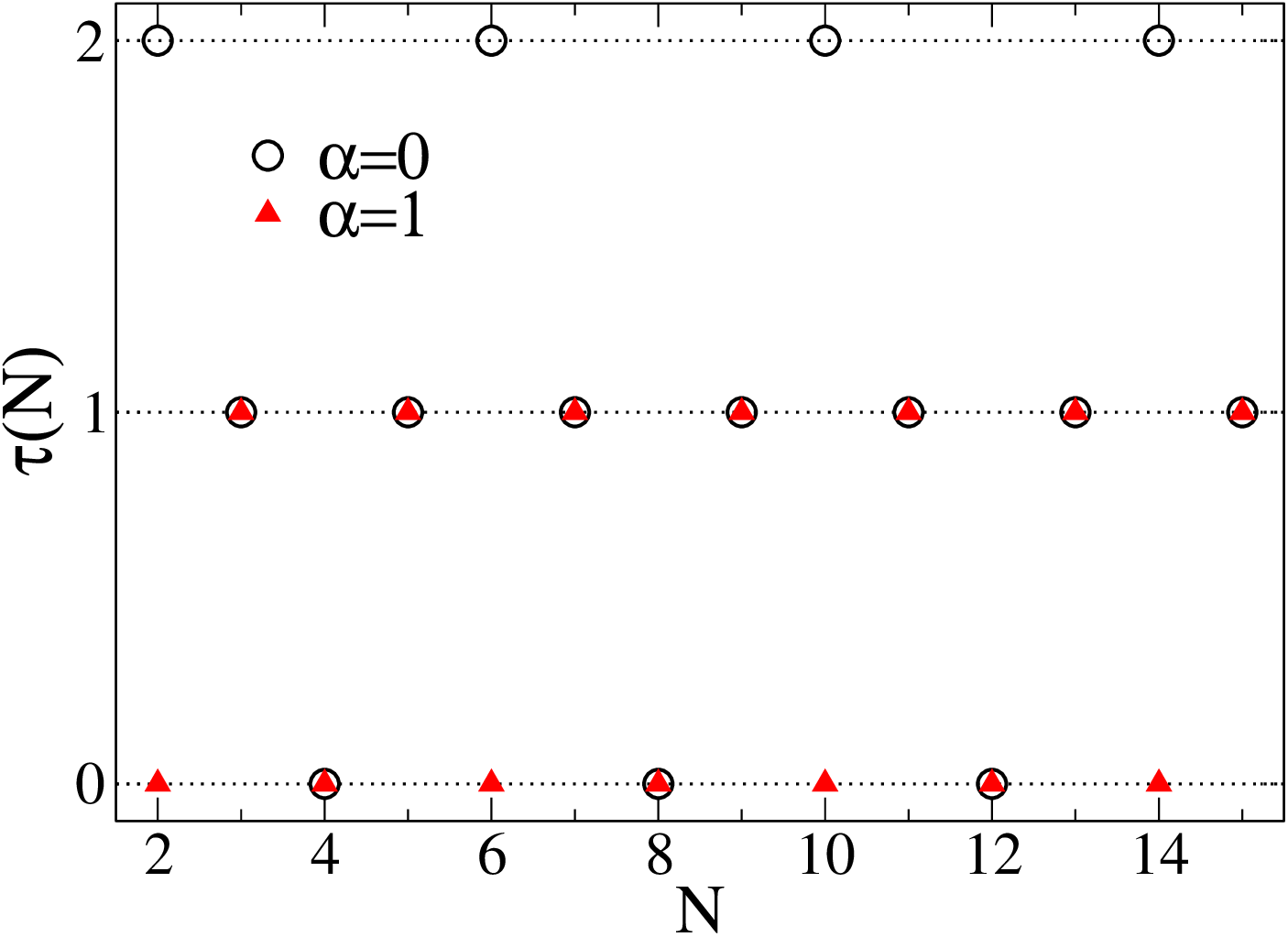}
\caption{\label{fig7} (Color online)
HF results for the ground-state valley polarization
$\tau(N)=\sum_{i=1}^N \tau_i$ in the zero-field case with $\alpha=0$
and $\alpha=1$. For $\alpha=0.5$, the same result as for $\alpha=1$
is found.\\ }
\end{center}
\end{figure}
\begin{figure}
\begin{center}
\includegraphics[width=3in]{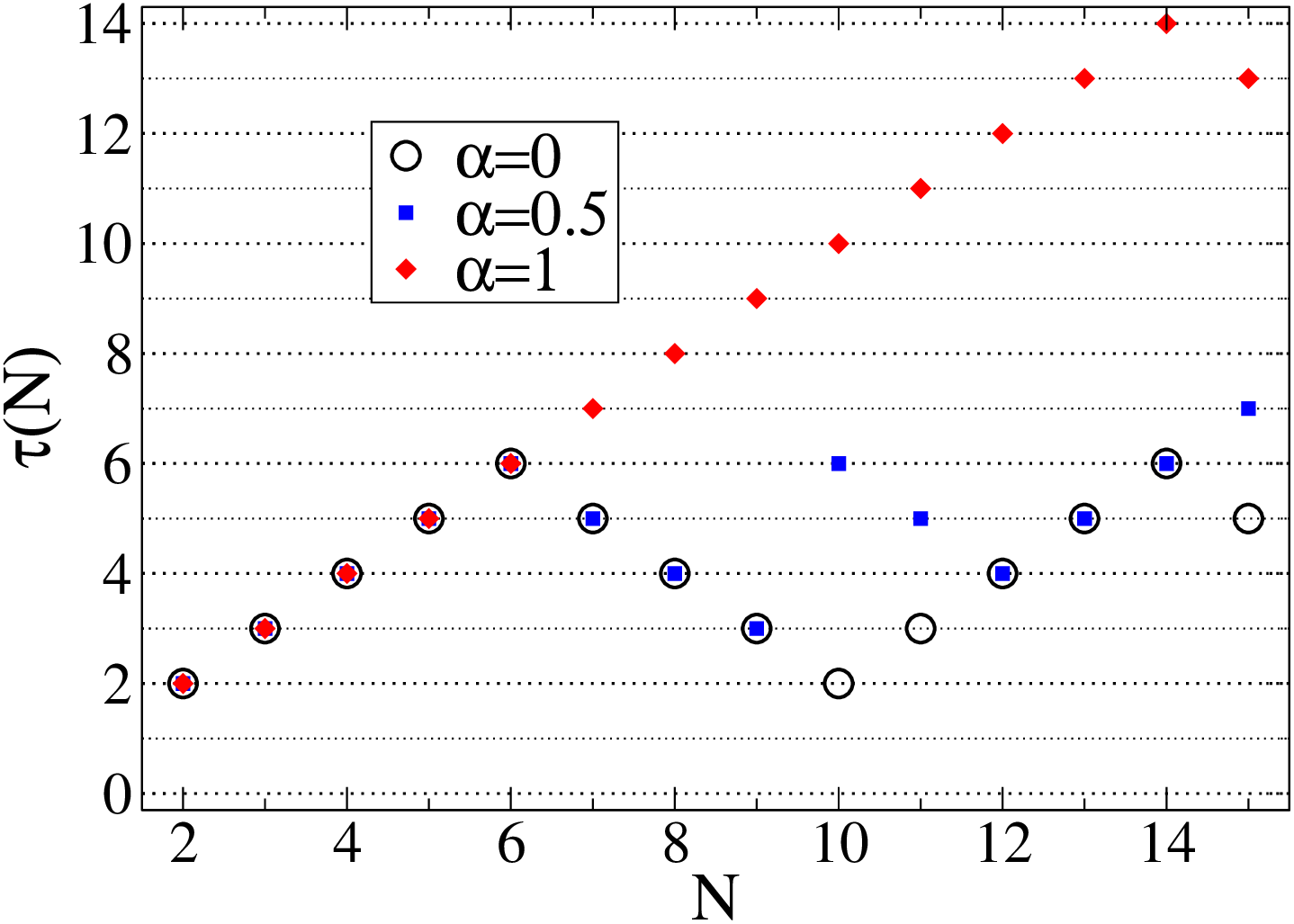}
\caption{\label{fig8} (Color online)
Same as Fig.~\ref{fig7} but for a perpendicular magnetic
field with $B=3$~T (with $R=30$~nm).\\  }
\end{center}
\end{figure}
\begin{figure}
\begin{center}
\includegraphics[width=3in]{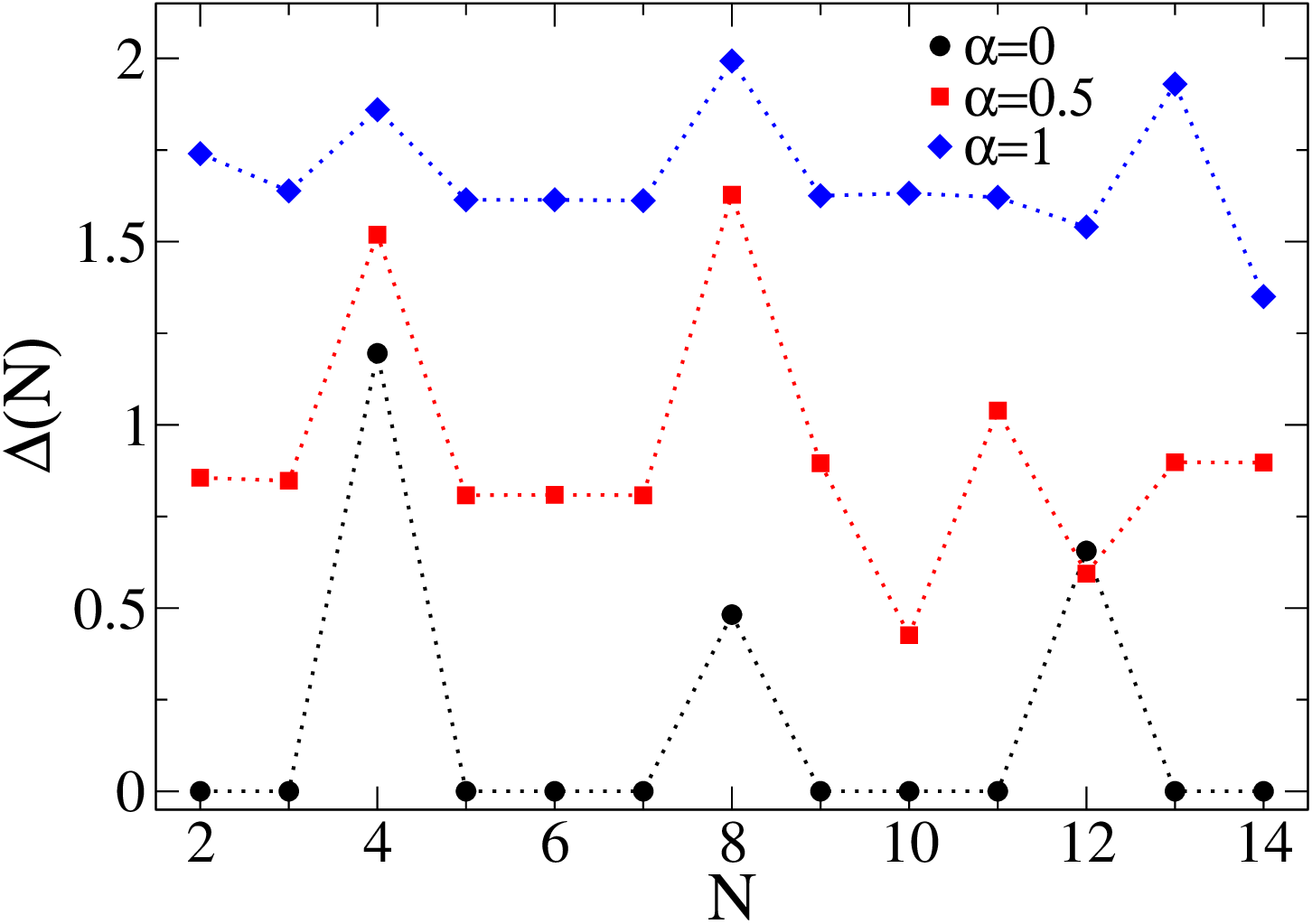}
\caption{\label{fig9} (Color online)
Addition energy \eqref{add} in units of $\Delta_0$ 
vs $N$ for several $\alpha$ from HF calculations for $H_+$ with $B=0$.\\ }
\end{center}
\end{figure}
\begin{figure}
\begin{center}
\includegraphics[width=3in]{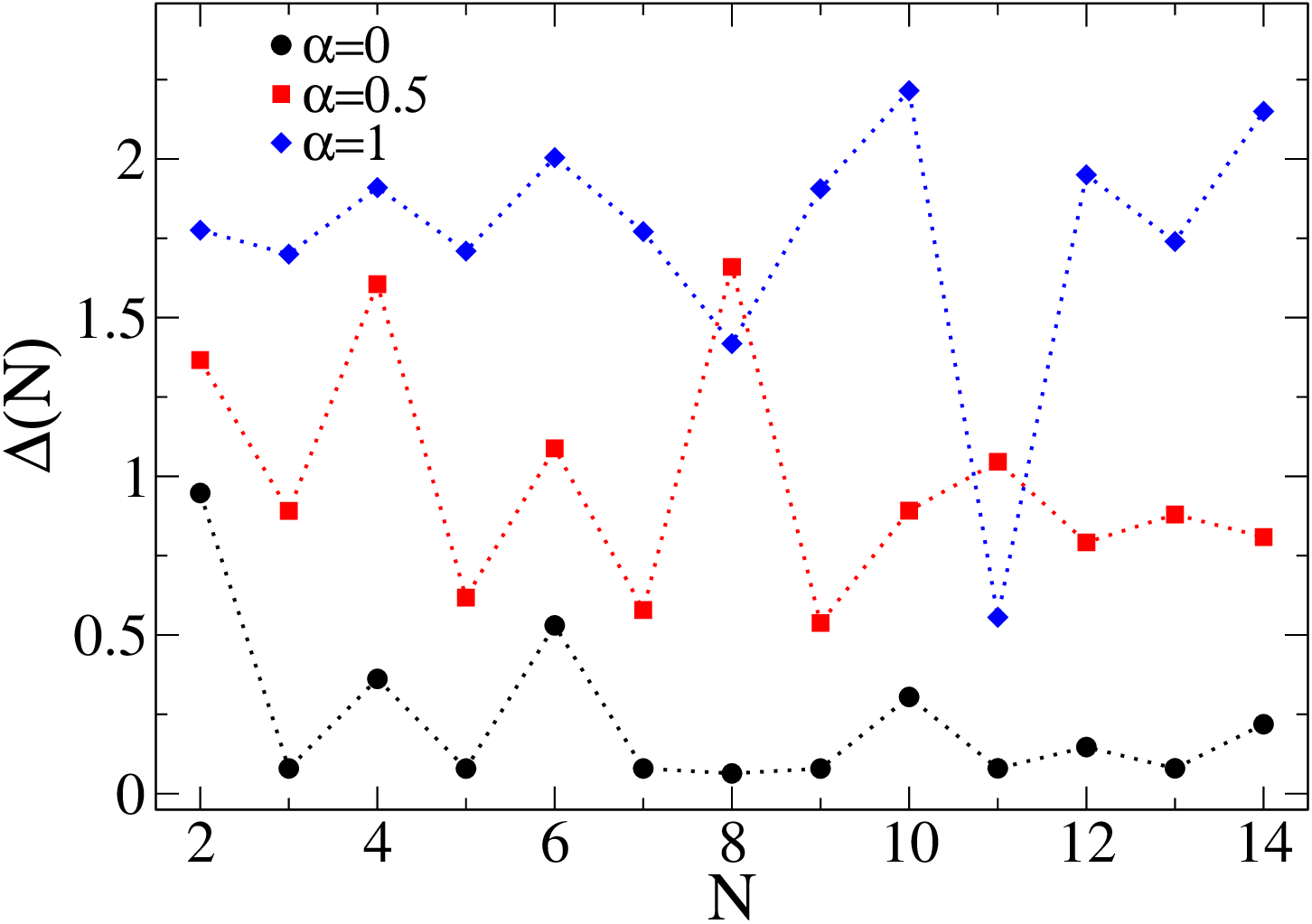}
\caption{\label{fig10} (Color online)
Same as Fig.~\ref{fig9} but with $B_x=15$~T and $B_z=3$~T (for $R=30$~nm).\\ }
\end{center}
\end{figure}

 Let us now turn to HF results for the ground state
of the $N$-electron dot with $N\le 15$ using the no-pair 
Hamiltonian $H_+$ [Eq.~\eqref{hplus}].  As discussed in Sec.~\ref{sec3},
this approximation is reliable only for weak interactions, and we 
focus on the regime $\alpha\le 1$ below.  
The spin and valley degrees of freedom are 
fully included in our self-consistent HF calculations.  

The total ground-state spin $S$ follows from the eigenvalue
$\hbar^2 S(S+1)$ of the total squared spin operator and is shown as
a function of $N$ in Fig.~\ref{fig6}, both for $B=0$ and 
in the perpendicular magnetic field $B=3$~T.  
For $N\le 14$ and $\alpha\le 1$, the spin filling sequence $S(N)$ is 
independent of the interaction strength $\alpha$ and displays 
a four-periodicity for $B=0$.   For $B\ne 0$, this periodicity
is reduced to a two-periodicity since now spin degeneracy is broken.  
We note that the spin filling sequence can be 
measured experimentally by Coulomb blockade spectroscopy.\cite{exp5}

The total ground-state valley polarization, $\tau(N)$, is shown
in Fig.~\ref{fig7} for $B=0$, and
in Fig.~\ref{fig8} under the perpendicular field $B=3$~T.
When $B=0$, the full Hamiltonian is symmetric under $\tau\to -\tau$,
and we here show only the positive solution. However, a  finite 
orbital field $B$ lifts the valley degeneracy.  We observe from 
Fig.~\ref{fig7} that for $B=0$, interactions reduce the 
four-periodicity of $\tau(N)$ for $\alpha=0$ down to a two-periodicity. This
can be understood by noting that the interaction of particles in 
different valleys is typically weaker than the intra-valley interaction.
For $B\ne 0$, this implies pronounced
interaction effects on the valley polarization. Figure \ref{fig8} shows that
strong interactions force subsequent particles to be added into the 
same valley, thereby valley-polarizing the $N$-particle system.
Both the spin and valley filling sequences obtained by HF theory
have been independently confirmed by ED of the no-pair Hamiltonian 
for $N\le 4$ (data not shown).

To estimate the accuracy of the HF approximation for the no-pair model,
we have also determined the relative difference between the
 HF ($E_{\rm HF}$) and the ED ($E_{\rm ED}$) energy, 
\[
\delta (N)= \frac{E_{\rm HF}(N,\alpha) - E_{\rm ED}(N,\alpha)}{
E_{\rm ED}(N,\alpha) -E_{\rm ED}(N,0)}.
\]
In all studied cases ($N\le 4$), $\delta(N)$ was found to be rather small.  
For instance, even when taking the large value $\alpha=1.5$, 
we obtain $\delta(2)=0.107$, $\delta(3)=0.175$ 
and $\delta(4)=0.148$.  As long as the no-pair approach
stays valid, we conclude that HF theory yields quite accurate results.

Our HF results for the \textit{addition energy},\cite{reimann} which
follows from the ground-state energy $E(N)$ using the relation
\begin{equation}\label{add}
\Delta(N) = E(N+1)+E(N-1)-2E(N),
\end{equation}
are shown in Fig.~\ref{fig9} for $B=0$ and several $\alpha$.
(Similar HF results but for the spinless single-channel version
were discussed in Ref.~\onlinecite{tomi}.)
Peaks in $\Delta(N)$ signify especially stable dot configurations (magic
numbers).  While for $\alpha=0$, $\Delta(N)$ again shows the
 four-periodicity due to  spin-valley degeneracy, 
the addition energy peaks become less pronounced with interactions,
and the four-periodicity is not always visible. 
Interestingly, while there are magic numbers $N=4,8,12,\ldots$ 
related to completely filled ``energy shells'' in the noninteracting case,
the addition energy curves $\Delta(N)$ are rather featureless and
almost flat for strong interactions.  This indicates that a 
constant interaction model\cite{cb} provides a reasonable
description, where the microscopic Coulomb interaction
is replaced by the electrostatic charging energy of an effective capacitor. 
The addition energy $\Delta(N)$ for $B\ne 0$ is shown in Fig.~\ref{fig10}.
The in-plane part $B_x$ of the magnetic field here acts
to increase the spin Zeeman field. 
However, Zeeman effects in graphene are weak,\cite{castro} and indeed
almost the same results as those in Fig.~\ref{fig10}
were found for $B_x=0$ and $B_z=3$~T.  As a consequence of the broken 
spin degeneracy, only an (approximate) two-periodicity in $\Delta(N)$ is 
observed in the magnetic field case.

\section{Conclusions}
\label{sec5}

In this work, we have studied the ground state properties of $N$ particles
in a disorder-free circular graphene quantum dot, with the filled Dirac
sea as the point of reference.  The boundary of the dot has been modelled
by the infinite-mass boundary condition, and the particles interact via 
the unscreened Coulomb potential whose prefactor is proportional to the 
bare dimensionless fine structure constant $\alpha$.  In contrast to
atomic physics where $\alpha=1/137$ is very small, in graphene
(e.g., by the variation of the substrate dielectric parameter) 
$\alpha$ may be tuned up to a maximum value of $\alpha\approx 2.2$ (reached for 
freely suspended samples).  For instance, a recent experiment\cite{exp5}
using Coulomb blockade spectroscopy for a graphene dot 
reported $\alpha\approx 1$.  We have studied the $N$-particle problem
in a graphene dot on various levels of complexity -- from the no-pair
Hamiltonian to the full QED model -- and by a number of different techniques.
Our main results are as follows.

By using exact diagonalization (ED) for $N=2$ and $3$ particles,
we found that the no-pair Hamiltonian $H_+$ originally proposed by 
Sucher,\cite{sucher} where the filled Dirac sea is assumed to be inert, 
is quantitatively reliable only for $\alpha\ll 1$, see
Sec.~\ref{sec3}.  While this represents the standard situation in 
atomic physics,\cite{reiher} 
it can easily be violated in graphene. For $\alpha\agt 0.5$,
our calculations indicate that electron-hole pair excitations contribute
to the ground state energy.  For $\alpha\agt 1$, these excitations 
proliferate and eventually cause a completely restructured ground state.
Technically, the projection operator $\Lambda_+$ defining the vacuum
should thus be changed to include interaction effects in a 
self-consistent manner. 
Mittleman\cite{mittleman} has shown that this goal can be achieved 
by first minimizing the ground state energy $E(N,\Lambda_+)$ 
for given $\Lambda_+$, followed by the maximization of the energy 
over all possible $\Lambda_+$.  The final result for $E(N)$
should then be equivalent to the QED results obtained numerically by ED
(in the limit of infinite basis size).

We here argue that graphene dots realize a finite-size crossover version of 
the bulk semimetal-insulator phase transition. We find that the crossover
scale is set by $\alpha\approx 1$, consistent with the bulk
result $\alpha_c\approx 1.1$.\cite{drut} 
When an orbital magnetic field is applied -- the Zeeman field plays no 
significant role --  electron-hole pair proliferation sets in 
earlier and implies a lowering of $\alpha_c$, 
consistent with the magnetic catalysis scenario.\cite{gusy}   
Even on a qualitative level, the no-pair Hamiltonian $H_+$ 
is thus reliable only on the semimetallic side of the transition
($\alpha\alt 1$).  

For the regime $\alpha\le 1$, we have reported detailed results 
using Hartree-Fock theory for $H_+$ and $N\le 15$ particles in 
Sec.~\ref{sec4}, taking into account the spin and valley degrees of freedom.
We find a four- (two-)periodicity in the spin filling 
sequence in the absence (presence) of a magnetic field, which can 
be understood from the single-particle picture and remains unaffected
by weak interactions.   However, the valley filling sequence is more 
intricate,  especially when $B\ne 0$. This is related to subtle 
differences between the intra- and inter-valley scattering matrix 
elements of the Coulomb interaction. We observe a strong tendency 
towards valley polarization induced by interactions in this $N$-body problem.
Finally, our analysis of the addition energy spectrum 
reveals that the constant interaction model can 
provide a reasonable description.

In our previous HF study of the spinless single-valley  no-pair
problem,\cite{tomi} we found that Wigner molecule formation sets
in for strong interactions.  Since that regime corresponds precisely
to the onset of electron-hole proliferation, $\alpha\agt 1$, where
the no-pair model becomes unreliable, we have analyzed the
question of Wigner molecule formation using ED for $N=3$ under
the full QED model again.  The Wigner molecule is identified from
pronounced density correlations, and our numerical results (not shown
here) are very similar to what we reported in Ref.~\onlinecite{tomi}.
We thus expect that the Wigner molecule formation is only weakly
affected by the electron-hole pair proliferation reported in this
paper.

\acknowledgments 

We thank A.~De~Martino, H.~Siedentop and E.~Stockmeyer for discussions.
This work was supported by the SFB TR 12 of the DFG.


\begin{thebibliography}{99}
\bibitem{foot1}
In the HF calculations, it is sometimes advantageous to
include a small mixing term in the single-particle Hamiltonian,
$H_{\rm mix} = \delta_K \tau_x+ \delta_s s_x.$
Inclusion of $H_{\rm mix}$ in the construction of the eigen-energies
and -states is straightforward. 
This allows us to probe all spin and valley states in one run, and by 
careful extrapolation $\delta_K,\delta_s\to 0$, we can extract the
ground state.
\end{thebibliography}
\end{document}